\RequirePackage{fix-cm}
\documentclass[12pt,twocolumn,epjc3]{svjour3}  
\smartqed  
\RequirePackage{graphicx}
\RequirePackage[colorlinks,citecolor=blue,urlcolor=blue,linkcolor=blue]{hyperref}
\journalname{Eur. Phys. J. C}
\begin{document}

\title{Mass and distance of AGN black holes from warped accretion disks
}



\author{A. Gonz\'alez-Ju\'arez\thanksref{e1,addr1}
        \and
        A. Herrera-Aguilar\thanksref{e2,addr1} 
}

\thankstext{e1}{e-mail: adrianag@ifuap.buap.mx}
\thankstext{e2}{e-mail: aherrera@ifuap.buap.mx}


\institute{Instituto de F\'isica, Benem\'erita Universidad Aut\'onoma de Puebla,\vspace{3pt}\\
Apdo. Postal J-48, CP 72570, Puebla, M\'exico \label{addr1}
}

\date{Received: date / Accepted: date}

\maketitle

\begin{abstract}
Along the last ten years, a general relativistic method has been developed to generate analytical expressions for the black hole (BH) parameters in terms of observables, namely the frequency shift of photons emitted by orbiting test particles and their positions on the sky. 
 Applications of the method to astrophysical systems such as Active Galactic Nuclei (AGNs), in particular to megamaser systems orbiting the central BH on their flat accretion disks, showed a coupling behavior in the mass-to-distance ratio $M/D$. Estimates for the ratio $M/D$ of a sample of BHs hosted at the core of several AGNs have been performed in recent years with the help of this method. However, both analytical expressions and statistical estimations depend only on the $M/D$ ratio rather than on independent parameters.
It is of current general interest to work with decoupled parameters in order to safeguard the intrinsic physical information encoded in each of them, given their high scientific relevance in understanding the structure of our Universe.
The purpose of this work is to find analytical expressions for the mass and distance of a Schwarzschild BH in terms of astrophysical observables by introducing a slight warping in the accretion disk of the orbiting megamasers.
As a result, independent analytical formulas for the mass and distance of AGN supermassive BHs are presented in terms of astrophysical observables: maser frequency shifts, disk parameters, and the galaxy's peculiar redshift.
\end{abstract}

\section{Introduction}
\label{intro}

The characterization of BHs and other astrophysical objects--encompassing the measurement of their mass, spin, distance from Earth, and recession velocity, among other parameters--continues to be a central focus of contemporary astronomical and astrophysical research.
Of particular scientific relevance is the development of new methods or techniques for determining the masses and distances of astrophysical sources, given the technical challenges involved. The past few decades have seen various contributions to mass and distance determination methods. 
Precise mass determination of neutron stars in binary systems has been performed using radio observations of rotation-powered pulsars \cite{Ozel2016}. Reverberation mapping has been employed to infer the mass of supermassive black holes (SMBHs) in situations when angular resolution is limited but adequate temporal resolution is available \cite{Peterson2014}. Another approach to estimating the masses of SMBHs located at the core of active galactic nuclei (AGNs) has been developed in \cite{Miyoshi1995}
(see the interesting review \cite{Moran2008}), and further advanced by the Megamaser Cosmology Project (MCP), where Keplerian rotation curves of water megamasers\footnote{Megamaser systems are astrophysical  configurations living at the core of AGNs and consisting of central supermassive BHs orbited by accretion disks that host water vapour clouds emitting at 22 GHz by stimulated emission with considerable luminosity.} 
orbiting AGN central sources have been statistically fitted 
\cite{MCPII,MCPIII,MCPIV,MCPV,MCPVI,MCPVII,MCPVIII,MCPIX,MCPX,MCPXI,MCPXII,MCPXIII} (see as well the comprehensive review \cite{Lo2005}). In our own galaxy, the central SMBH mass, based on the dynamics of the nearby stellar cluster around its core, has been  estimated  \cite{AndreaG1998,Gillesen2009}.
On the other hand, significant progress in measuring distances to astrophysical bodies has been made by integrating geometric and traditional methods (as the use of standard candles) to construct scaling relations \cite{Czerny2018}; examples include modern geometric parallax, which now offers longer baselines \cite{Gisela2017}, star clusters instead of single objects and the use of star spectra instead of changes in apparent position of the stars \cite{Gossan2012}
(see also the review \cite{Tevenin2017}). In parallel, the development of suitable statistical methods for AGNs has made it possible to estimate their distances with remarkably high precision \cite{MCPII,MCPIV,MCPV,MCPVI,MCPVIII,MCPIX,MCPXI,MCPXIII,Argon2007,Humphreys2008,Humphreys2013}.

The difficulties addressed in determining the masses of astrophysical objects starts realizing that they can only be determined through indirect methods. This task is further complicated when addressing black hole masses due to their intrinsic physical properties: no signal coming from the interior of the event horizon can be detected by a distant observer and, therefore, these compact objects cannot be directly seen. As a result, indirect dynamical methods that use nearby moving objects or photometric treatments for mass determination must be implemented.

Mirroring the challenges of mass determination, similar difficulties arise when inferring the distances of astrophysical objects since this problem must be addressed at different scales (the so-called scale ladder). First of all, there is no single method that works for all scales and covers the total required distance to be measured, as a consequence, the uncertainties exacerbate with each step up
\cite{Czerny2018,Gossan2012}. Another important \textcolor{blue}{
def\/iance} is that celestial objects are located in a three-dimensional space while their observation occurs in two, demanding the development of necessary methods to add the third dimension accurately and without introducing false information, or, conversely, to adapt physical models that even in two dimensions are capable of explaining what is observed. 
Finally, several properties of astrophysical objects can be determined in terms of their distance, such as brightness and magnitudes (through the distance modulus) and size (through the quotient of distance and angular diameter), to name a few.

Regarding the BH mass and distance determination in megamasers systems, in addition to the aforemetnioned proven efficacy of post-Newtonian approaches, a general relativistic formalism is likely to supply a more comprehensive and precise description of gravitational phenomena in the strong field regime, even for a wider gamma of astrophysical configurations. Following this line of thought, the detection of the gravitational redshift \cite{Grav18,Do2019} and the Schwarzschild precession \cite{Grav20} in the motion of the S2 star orbiting our SMBH SgrA$^*$, the imaging of the BH shadow originating at the event horizon vicinity of both M87$^*$ \cite{EHT1} and SgrA$^*$ \cite{EHT2}, as well as the recently reported evidence for the Bardeen-Petterson effect in the accretion disk of M87$^*$ \cite{Iorio25} vividly encourage the design and implementation of relativistic models for studying the physics around SMBHs.

Within this general relativistic framework, an analytical method that expresses the mass and spin parameters of a Kerr BH as a function of astrophysical observational data was presented in \cite{Herrera15,Banerjee22} as well as the parameters of more general spherical and axisymmetric BHs \cite{Sharif2016,Kraniotis2021,Lopez2021,Giambo2022,Morales2024}. 
A further development of the method framed in the Kerr-de Sitter spacetime rendered closed formulas for both the BH parameters and the Hubble constant in terms of observational data \cite{Momennia2023}, a further study estimated these quantities within the Schwarzschild-de Sitter spacetime \cite{Villaraos2025}.
In addition, the method has been implemented to analytically express the BH parameters of several modified gravity theories in the language of astrophysical observables \cite{Sheoran2018,Shankar2018,Uniyal2018,Debnath2021,Mustafa2022,Fu2023,Alibekov2023,Saidov2024,Martinez2024} as well as to statistically estimate the parameters of conformal gravity BHs \cite{Martinez2025}.

Thus, within general relativity the estimation of the principal BH parameters of several megamaser systems hosted at the core of AGNs has been performed in \cite{Nucamendi21,Artemisa22,Villaraos,Adriana2024}. In these studies, the authors express the frequency shift in terms of central BH mass-to-distance ratios and sky positions of photon sources dwelving on their accretion disks. By expanding this analytical formula with respect to the ratio $M/r_e\ll 1$, where $M$ is the BH mass and $r_e$ denotes the maser orbital radius, the leading general relativistic correction corresponds
to the gravitational redshift and reads $3M/(2r_e)$ (for further details please refer to our review \cite{Gonzalez2025} and references therein). The magnitude of this correction takes different values of the same order in several megamaser systems; for example, for the closest maser to the BH in NGC 4258 the gravitational redshift amounts to $7.45$ km s$^{-1}$ \cite{Nucamendi21}. A bit smaller magnitudes for the gravitational redshift for over a dozen of AGN BHs have been reported in \cite{Villaraos,Adriana2024}, evidencing that we are at the threshold of detecting this curvature effect that has been underestimated so
far in the literature.\\
Thus, the works \cite{Herrera15,Banerjee22} constitute the theoretical basis of the above referred general relativistic method. The required observables are the frequency shift of photons emitted by orbiting test particles and their projected position, i.e. their Right Ascension and Declination, on the plane of the sky. 
It should be noted that this method allows for a clear identification and quantification of general and special relativistic effects \cite{Nucamendi21} such as the gravitational redshift produced by the spacetime curvature generated by the BH and the peculiar redshift originated by the recessional motion of the host galaxy, respectively.
It is worth mentioning that the frequency shift is a general relativistic invariant by definition. Remarkably, by appropriately taking into account another general relativistic invariant observable, namely, the redshift rapidity (the derivative of the frequency shift with respect to the proper time), the method allows for separately expressing both the mass and distance of the BH in terms of purely observable astrophysical quantities \cite{Momennia2024}. Moreover, this decoupling procedure avoids the need to take external experiments into account. However, the measurement of this quantity at the positions lying around the midline of the accretion disk is not an easy task, because of its negligible magnitude and the resolution needed to precisely track the trajectory of a given maser in that region. The situation is different for systemic masers lying close to the line of sight (LOS), where the measurement of the redshift change over time is maximum, in windows of a couple of years, and technically more feasible  \cite{MCPII,MCPIV,MCPV,MCPVI,MCPVII,MCPVIII,MCPXI,MCPXIII,Humphreys2008,Yamauchi2005,Baan2022,Braatz2024}.
\\
The inclusion of the warped disk model into BH accretion disks studies follows from the observation of a mildly misaligned maser feature distribution on the sky in very long baseline interferometry (VLBI) maps (see \cite{Humphreys2008,Humphreys2013,Herrnstein05}). Thus, the disk warping is a necessary observable to be taken into account in existing accretion disk models of AGN megamaser systems. In this context, slightly deformed disk models have been considered when fitting the BH parameters in several works \cite{MCPIV,MCPV,MCPVIII,MCPX,MCPXI,MCPXII,Herrnstein05}, a fact that consequently has allowed for statistical estimations of distances to the corresponding central BHs from Earth. Typically, a slight scattering in the maser map along the Right Ascension axis on the plane of the sky is observed. Therefore, for modeling the accretion disk warping, slight changes that depend on the emitter radius in the observed inclination angle are implemented.
Thus, in this work, a new decoupling procedure that incorporates a slight warping of the accretion disk in the expression for the frequency shift is presented. Once the disk warping has been introduced in the frequency shift, analytical expressions for the mass and distance parameters of the BH are naturally obtained in terms of astrophysical observables, disk parameters and the peculiar velocity of the host galaxy. It is worth noticing that this decoupling can only take place within general relativity due to the asymmetric nature of the total frequency shift expression. In contrast, in the Newtonian approach it is impossible to decouple these quantities from the blueshift and redshift expressions because these frequency displacements are completely symmetric and cannot be separated without an independent equation.\\
The organization of the paper is as follows. In Section \ref{sec:2} a brief description of the general relativistic me\-thod for calculating the mass-to-distance ratio of supermassive BHs at the core of AGNs is made. Section \ref{ss2.1} presents the details of the formalism used to introduce and parameterize a warped accretion disk in the setup achieved by introducing a slight disk deformation into the analytical expression for the frequency shift of photons emitted by maser features, while Subsection \ref{ss2.2} displays the decoupling procedure for setting the mass and distance of the BH in terms of astrophysical observables, disk parameters and peculiar motion of the host galaxy. Finally, in Section \ref{sec:3} a discussion about the obtained results and their astrophysical implications is developed.
\section{GR method and warping of the accretion disk}
\label{sec:2}
The general relativistic (GR) model for BH rotation curves implemented in this work can be used to study the cores of AGNs, in particular the BHs orbited by maser clouds in their accretion disks. 
The method links the effect of the spacetime curvature generated by a central compact object on the geodesic motion of nearby massive and massless particles with the frequency shift of the photons emitted by the orbiting bodies and detected on Earth.

The method also allows for obtaining analytical expressions for the parameters of the BH, in the case of equatorial circular orbits for the emitter and a static distant observer with respect to the BH, in terms of astrophysical observables: the redshift and the blueshift of photons emitted by the aforementioned masers and their orbital positions on the sky. 
In addition, in order to perform BH parameter estimations, it is possible to implement a Bayesian statistical treatment to fit astrophysical megamaser observational data.\\

In this work the Schwarzschild metric is considered 
\begin{equation}
ds^{2}=\frac{dr^{2}}{f}+r^{2}(d\theta ^{2}+\sin^2 {\theta }d\varphi
^{2})-fdt^{2},\quad f=1-\frac{2M}{r}.  \label{schw}
\end{equation}
Here $M$ is the total mass of the BH and geometrized units ($G=1=c$) are being used.

The frequency is a general relativistic invariant defined as 
$$
\omega=-k_{\mu}U^{\mu},
$$ 
where $U^{\mu}$ is the four-velocity vector that characterizes the geodesic orbital motion of test massive particles, $k^{\mu}$ is the four-momentum ($\hbar=1$) of the photons along their geodesic path and index $\mu=0,1,2,3$. 
Using this definition in the points of emission and detection, the expression for the frequency shift of the photons emitted by megamaser features in the equatorial plane ($\theta=\pi/2$ and $U^\theta=0=k^\theta$) of the Schwarzschild background and detected on Earth reads
\begin{eqnarray}
1 + z_{Schw} = \frac{\omega_e}{\omega_d} &=& \frac{\left(E_{\gamma} U^t - L_{\gamma} U^{\varphi} - g_{rr} k^rU^r \right)_e }{\left(E_{\gamma} U^t - L_{\gamma} U^{\varphi} - g_{rr} k^rU^r \right)_d},
\label{zSchw1,2}
\end{eqnarray}
here indices $_{e}$ and $_{d}$ denote emission and detection points, $E_\gamma$ is the conserved energy and $L_\gamma$ is the preserved angular momentum  of massless particles. These quantities are defined as
\begin{eqnarray}
  E_{\gamma} &=& - g_{\mu\nu} \xi^{\mu} k^{\nu} = - g_{tt}k^{t} , \\
  L_{\gamma} &=&  g_{\mu\nu} \psi^{\mu} k^{\nu} = g_{\varphi\varphi} k^{\varphi}, \qquad 
\end{eqnarray}
where $ g_{\mu\nu}$ is the metric tensor, whereas $\xi^\mu=(1,0,0,0)=\delta_t^\mu$ and $\psi^\mu=(0,0,0,1)=\delta_{\varphi}^\mu$ are the temporal and rotational Killing vector fields, respectively. 

We further consider that emitters are in geodesic circular motion around the BH (meaning $U^r=0$ for the maser clouds) on a flat thin disk.
Additionally, we particularize the light bending parameter\footnote{Also known in Astronomy as the impact parameter.} on either side of the LOS at the midline (where $k^r=0$) as
\begin{eqnarray}
b_{\pm} \equiv \frac{L_{\gamma}}{E_{\gamma}} &=&  \pm \sqrt{-\frac{g_{\varphi\varphi}}{g_{tt}}} 
= \pm \frac{r_e}{\sqrt{1 - 2\tilde{M}}},
\end{eqnarray}
where $\tilde{M}=M/r_e$, $r_e\approx D\Theta$ is the orbital radius of a given maser feature located on the midline (at $\varphi=\pm\pi/2$), $D$ stands for the BH distance from Earth, and 
$$
\Theta=\sqrt{ \left( x_{i}-x_{0}\right) ^{2}+\left( y_{i}-y_{0}\right) ^{2}}
$$ 
with ($x_i$, $y_i)$ denoting the position of the $i$-th megamaser on the sky and ($x_0$, $y_0)$ is the BH position.

Thus, by considering a static detector ($U^\varphi_d=0=U^r_d$) located far away from the BH, i.e. when $r_d\longrightarrow \infty$, its four-velocity simplifies to $\left.U^{\mu}\right|_d = (1,0,0,0)$ rendering the following total frequency shift
\begin{eqnarray}
1 + z_{_{Schw_{1,2}}} &=& 
\left( U^t - b_{\mp} U^{\varphi} \right)_e 
\nonumber\\
&=&\frac{1}{\sqrt{1-3\tilde{M}}}\left(1\pm\frac{\sqrt{\tilde{M}}}{\sqrt{1-2\tilde{M}}} \right),
\label{zSchw1,2}
\end{eqnarray}
where the indices $_{1,2}$ indicate the corresponding redshift and blueshift of photons. These are precisely the physical quantities that are measured in megamaser systems at the core of AGN with the aid of VLBI techniques, and reported as velocities with the help of the optical definition $z=v/c$, (see for instance \cite{MCPIII,MCPIX}). 

It is worth noticing that this observable depends just on the $\tilde{M}=M/r_e\approx M/D\Theta$ ratio and therefore the $M$ and $D$ parameters are intrinsically coupled.

A closed formula for the BH $M/D$ ratio in the language of frequency shifts can be extracted from relations (\ref{zSchw1,2}) as follows \cite{Banerjee22}
\begin{equation}
    \frac{M}{D}\approx\tilde{M}\Theta
    =\frac{\Theta}{3}\left[\frac{(R_S+B_S)^2-4}{(R_S+B_S)^2}\right],
    \label{eq:Mtilde0}
\end{equation}
where the notation 
$$
1+z_{Schw_{1}}\equiv R_S, \qquad
1+z_{Schw_{2}}\equiv B_S
$$
has been introduced.

\subsection{Peculiar motion of the BH host galaxy}

Galaxies are moving with respect to each other in our Universe, in particular, with respect to an observer located on Earth.

The receding or approaching local motion of a given AGN with respect to Earth with constant peculiar velocity $v_p$ (which produces a peculiar redshift $z_p$) can be taken into account by considering a special relativistic boost \cite{RindlerSR1989}
\begin{equation}
1 + z_{boost} = \gamma (1 + \beta \cos\kappa), \qquad \gamma=1/\sqrt{1-\beta^2}, 
\label{z_boost}  
\end{equation}
where $\beta=v_p/c\equiv z_p$, $c$ is the speed of light and $\kappa$ is the angle between the galaxy's vector velocity and the LOS. From now on this angle is considered to adopt the value $\kappa=0$ to account for the recessional motion of the galaxy from a distant observer in the radial direction.

Therefore, the total frequency shift results as the composition of the Schwarzschild redshift and the peculiar frequency shift \cite{Davis14}
\begin{equation}
    1+z_{tot}=(1+z_{_{Schw}})(1+z_{boost}),
    \label{eq:composition}
\end{equation}
yielding the following expression for the total redshift 
\begin{eqnarray}
    1+z_{tot_{1,2}}&\!=\!&(1+z_g+z_{kin_\pm})(1+z_{boost})\\
    &\!=\!&\frac{1}{\sqrt{1-3\tilde{M}}}\left(1\pm\frac{\sqrt{\tilde{M}}}{\sqrt{1-2\tilde{M}}} \right)\!
    \sqrt{\frac{1+z_p}{1-z_p}}.\nonumber
    \label{eq:R0}
\end{eqnarray}
which encompasses the gravitational redshift $z_g$ generated by the BH curvature of spacetime, the kinematic frequency shift $z_{kin_\pm}$ produced by the orbital motion of the test particles around the BH, and the peculiar redshift $z_p$ that describes the local peculiar motion of the AGN relative to the Earth.

This relation will be used further when decoupling the expressions for the mass and distance parameters of a BH and setting them in the language of astrophysical observational quantities when the accretion disk warping will be introduced.

\section{Modeling the disk warping 
}
\label{ss2.1}

Now the incorporation of the warping of the accretion disk into the expression for the redshift to account for the maser distribution map that is observed on the sky needs to be considered.

In order to introduce this deformation, slight changes depending on the emitter radius in the observed inclination angle require to be implemented. However, these small orbital deviations of the masers from the equatorial plane are such that both $U^r$ and $U^\theta$ should remain practically zero.
 
\begin{figure*}
\centering
\includegraphics[width=.6\textwidth]{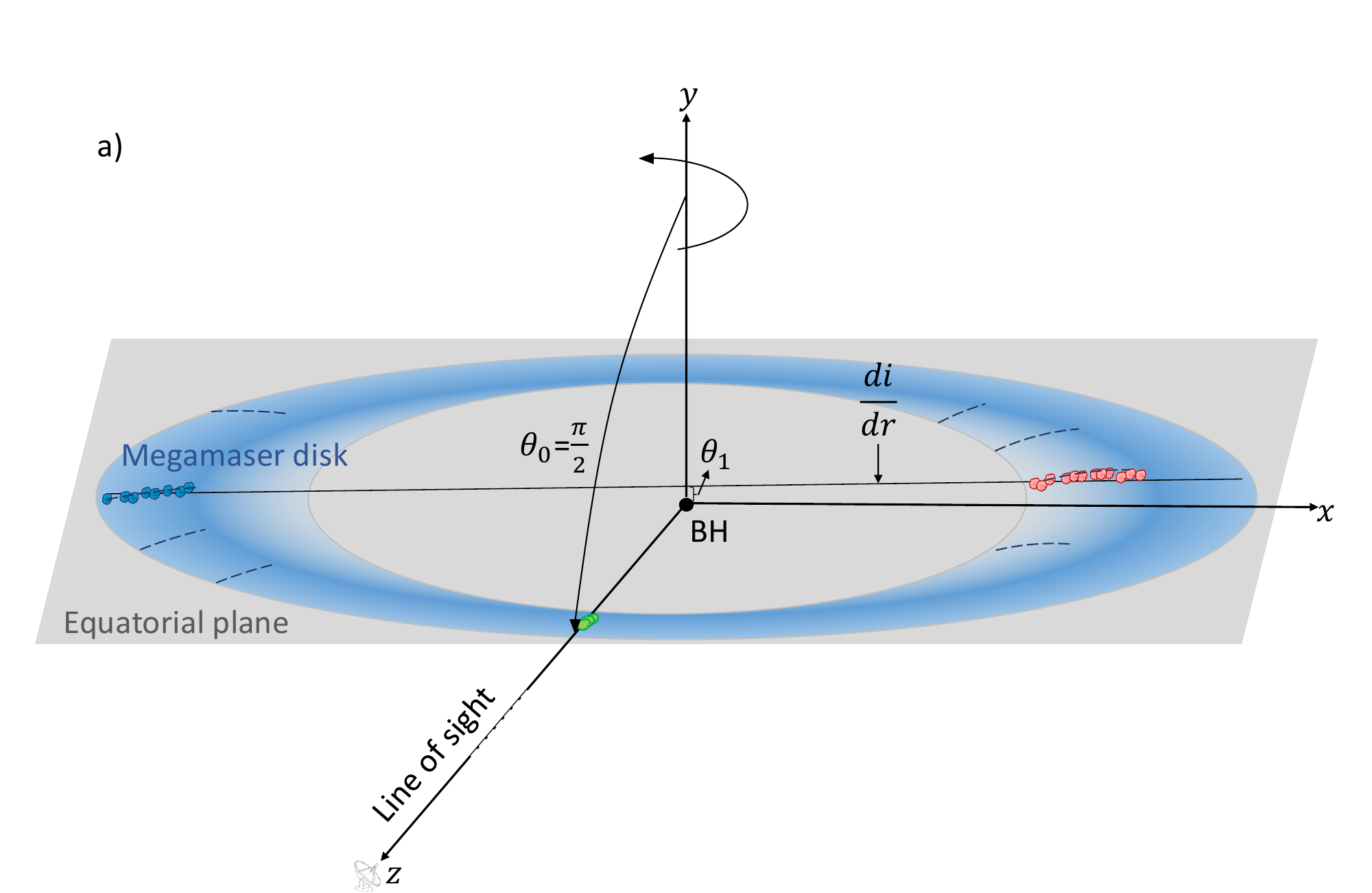}\\
\includegraphics[width=.6\textwidth]{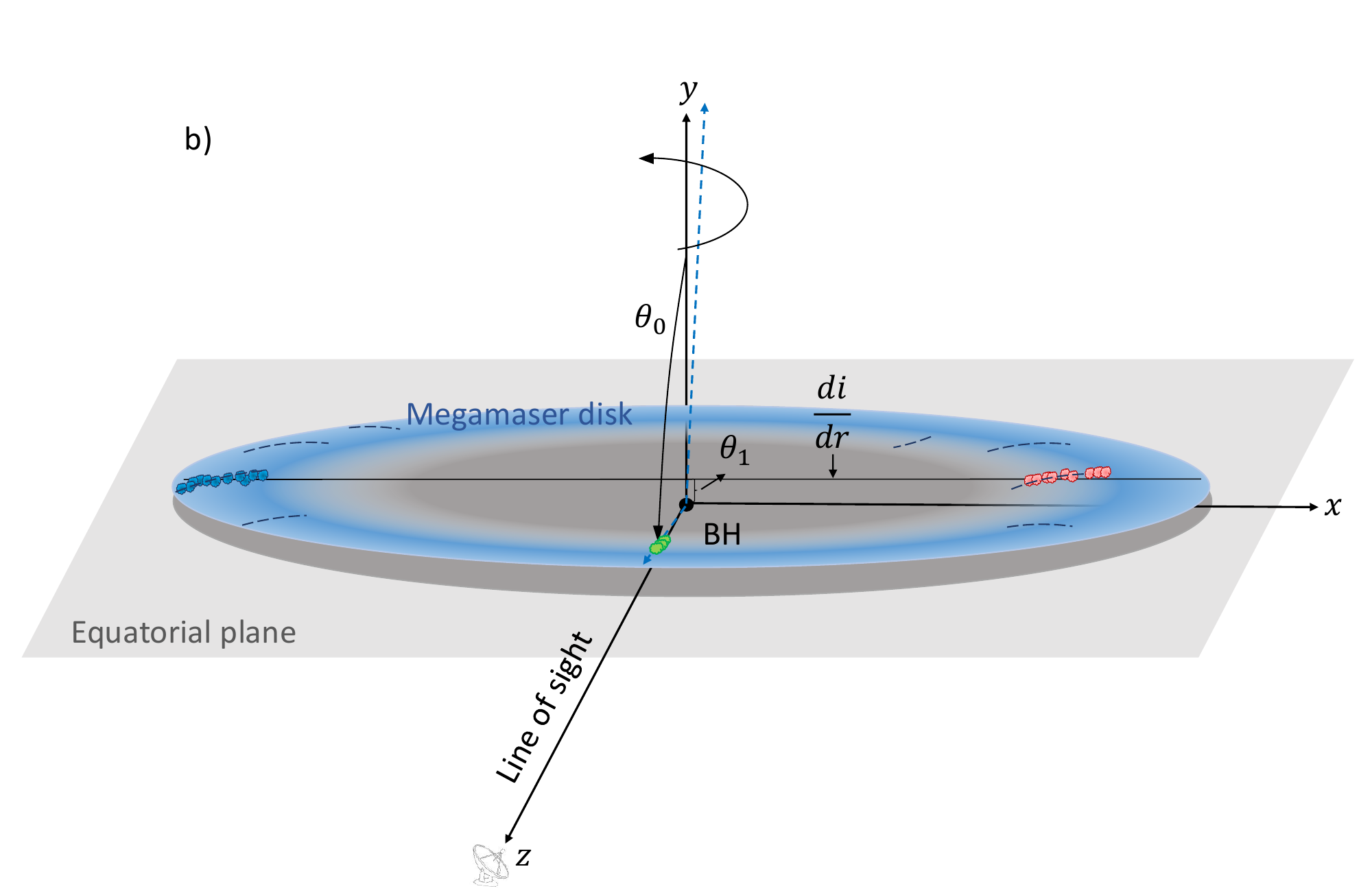}
\caption{
Schematic diagram of the geometry of a linearly warped disk hosting a megamaser system revolving around a black hole. The maser disk is shown in blue lying on the equatorial plane ($\theta_0=\pi/2$) in a) and with a slight upward tilt relative to the equatorial plane in b). Here, slight deformations of the disk are represented by dashed blue lines. The sky distribution of the highly redshifted and blueshifted masers features illustrates that the disk deformation is modeled in such a way that all masing spots approximately lie on a single straight line which crosses the polar axis at angle $\theta_1$ and has a slope $di/dr$ as measured by a distant observer for a coordinate system anchored in the BH. This linear disk warp is the same for any azimuthal direction in our model. This representation is the simplest possible linear disk warping seen by a far away observer, and is the one described by our frequency shift model (the one described by Eqs. (\ref{eq:R})-(\ref{eq:x}). However, different maser configurations can arise depending on the way the disk is geometrically deformed.}
\label{fig:WarpDiskP}
\end{figure*}

In contrast to the work \cite{Herrnstein05} (see also the following research papers by the MCP collaboration \cite{MCPIV,MCPV,MCPVIII,MCPX,MCPXI,MCPXII}) where a linear deformation of the accretion disk is introduced into the inclination and position angles of their model, here a warped configuration just on the equatorial plane is considered, a treatment that takes advantage of spacetime spherical symmetry. Then, by considering that the BH accretion disk is tilted relative to the equatorial plane\footnote{This procedure is mathematically parameterized by projecting the disk to the equatorial plane through a dot product of the LOS and the axis that joins the BH with the accretion disk at the points on which the azimuthal angle vanishes, yielding the factor $\cos(\pi/2-\theta_0)=\sin\theta_0$ in the kinetic term of the total redshift.} with an inclination angle $\theta_0\ne\pi/2$ (see Fig. \ref{fig:WarpDiskP}), the corresponding expression for the total redshift reads 
\begin{eqnarray}
1+z_{tot}=(1+z_g+z_{kin_{_\pm}}\sin{\theta_0})(1+z_{boost}).\nonumber
\end{eqnarray}
By further allowing for a slight disk warping as a first approximation, 
the following expression is obtained
\begin{eqnarray}
    1+z_{tot_{1,2}}&\!=\!&\frac{1}{\sqrt{1-3\tilde{M}}}\left(1\pm\frac{\sqrt{\tilde{M}}\sin{x}}{\sqrt{1-2\tilde{M}}} \right)\!
    \sqrt{\frac{1+z_p}{1-z_p}},
    \label{eq:R}
\end{eqnarray}
where the simplest sinus argument (linear in emitter radius in the inclination angle) reads \begin{equation}
x=\theta_0+\tilde\theta_1+\frac{di}{dr}r_e=
\theta_0+D\left(k\,\theta_1+\frac{di}{dr}\Theta\right);
\label{eq:x}
\end{equation}
here $\theta_0$ is usually fixed with the aid of maser map observational data; $\tilde\theta_1=k\,\theta_1$, where $k$ is a unit constant with inverse length dimension, $\theta_1$ is the intercept of the linear warping with the Declination axis; $di/dr$ being the slope or radial gradient of inclination of the deformed accretion disk with respect to the equatorial plane, it also has inverse length dimension; and the approximation $r_e\approx D\Theta$ has been employed. In principle, the parameters $\theta_1$ and $di/dr$ can also be determined from the maser map on the sky. A diagram illustrating the above description is given in Fig. \ref{fig:WarpDiskP}.

\noindent
It is worth mentioning that it is precisely the introduction of the disk warping, that encodes information about the BH distance to Earth, in the $\sin x$ function (\ref{eq:x}) of the relativistic black hole rotation curve model (\ref{eq:R}) that breaks the degeneracy between $M$ and $D$. In contrast, in the Keplerian model these parameters are necessary coupled through the product $\sqrt{\tilde{M}}\sin x$, preventing the separation of $M$ and $D$. Notwithstanding, the three parameters $\theta_0$, $\theta_1$ and $di/dr$ that enter the $\sin x$ function produce a degeneracy, since multiple sets of their values can generate the same observable output.

\subsection{BH mass and distance in terms of observables}
\label{ss2.2}

Since the warping in the accretion disk introduces a sole dependency on the BH distance $D$, it breaks the degeneracy observed in the BH mass-to-distance ratio of the equatorial redshift (\ref{eq:R0}) for the maser features located along the midline (the masers with azimuthal angle $\varphi=\pm \pi/2$), remarkably enabling the decoupling of these quantities.

Now, in order to simplify further calculations, the notation 
\begin{equation}
1+z_{tot_{1}}\equiv R, \qquad
1+z_{tot_{2}}\equiv B
\label{eq:RandB}
\end{equation}
will be used onward.
Thus, by calculating the sum $R+B$ from Eq. (\ref{eq:R}), 
an expression for the ratio $M/D$ is obtained
\begin{equation}
    \frac{M}{D}\approx\tilde{M}\Theta
    =\frac{\Theta}
    {3}\left[\frac{(1-z_p)(R+B)^2-4(1+z_p)}{(1-z_p)(R+B)^2}\right].
    \label{eq:Mtilde}
\end{equation}
Then, by computing the subtraction $R-B$, making further use of the convenient combination
\begin{equation}
    \left(\frac{R-B}{R+B}\right)^2=\frac{\tilde{M}\sin^2{x}}{1-2\tilde{M}},
    \label{eq:RBcombine}
\end{equation}
and solving for $\sin{x}$, an expression for the distance $D$ is isolated
\begin{eqnarray}
    &&D=\frac{1}{\left(k\,\theta_1+\frac{di}{dr}\Theta\right)}\times
    \label{eq:Dsolita}\\
&&\left[\!\sin\!^{-1}\!\!\left(\!\sqrt{
    \frac{(1\!-\!z_p)(R+B)^2+8(1+z_p)}{(1-z_p)(R\!+\!B)^2-4(1+z_p)}}\frac{R-B}{R+B}\right)\!-\!\theta_0\right]\!.
    \nonumber
\end{eqnarray}
This is an analytical formula for the BH distance from an observer located on Earth in terms of the observational quantities $R$, $B$ and $\Theta$, the parameters of the accretion disk $\theta_0$, $\theta_1$ and $\frac{di}{dr}$, and the peculiar redshift $z_p$. 

It is worth to remark that according to the approximation for $r_e\approx D\Theta$ and using Eq. (\ref{eq:Dsolita}), an expression for the orbital radius of a given highly frequency shifted photon source, which is not an observable quantity, can also be obtained in terms of the same observable quantities and warped disk parameters as the distance $D$.

Substituting the expression for the distance, Eq. (\ref{eq:Dsolita}), into Eq. (\ref{eq:Mtilde}) finally renders
\begin{eqnarray}
    &&M\!=\!\frac{\Theta}{3\left(k\,\theta_1\!+\!\frac{di}{dr}\Theta\right)}\!\!\left[\frac{(1-z_p)(R+B)^2-4(1+z_p)}{(1-z_p)(R+B)^2}\right]\!
    \times\nonumber\\ &&\left[\!\sin\!^{-1}\!\!\left(\!\sqrt{
    \frac{(1\!-\!z_p)(R+B)^2+8(1+z_p)}{(1-z_p)(R\!+\!B)^2-4(1+z_p)}}\frac{R-B}{R+B}\right)\!-\!\theta_0\right]\!,
    \nonumber
    \label{eq:Msolita}\\
\end{eqnarray}
a closed formula for the BH mass expressed again in terms of astrophysical observational data, disk parameters and the peculiar redshift of a given galaxy.

It is worth noting that with the aid of the H21 spectral line of hydrogen gas present in a given galaxy and the Hubble law, the peculiar redshift $z_p$ can be determined \cite{MCPVII}. On the other hand, by performing and appropriately correlating independent astrophysical observations when studying the maser disk map, the accretion disk parameters can in principle also be inferred.




\subsection{General relativistic vs Newtonian approach}
It is worth underlining that $M$ and $D$ decouple in the general relativistic framework thanks to the structure of the equations for the frequency shifts (\ref{eq:R}). This relation reflects the asymmetric nature of redshifts and blueshifts due to presence of the gravitational redshift in the first term that encodes information about the general relativistic corrections to the Newtonian expression through $\tilde{M}$. The second term is the kinematic frequency shift that describes the Newtonian orbital motion of the test particle around the BH and contains $D$ alone in the argument of the sinus function. This is precisely the fact that allows for the decoupling of the $M$ and $D$ parameters.

In contrast, the corresponding expression for the Newtonian rotational velocity (in the sense of maser features located around the midline) including the disk warping reads
\begin{eqnarray}
    v_{rot_{1,2}}=\pm\sqrt{\tilde{M}}\,\sin{x},
    \label{eq:RotNewtonVel}
\end{eqnarray}
therefore the total Newtonian velocity that also accounts for peculiar motion is
\begin{eqnarray}
    v_{tot_{1,2}}=\pm\sqrt{\tilde{M}}\,\sin{x}+v_p.
    \label{eq:NewtonVel}
\end{eqnarray}
As it can be seen from the latter equations, it do not decouple $M$ and $D$ in the Newtonian approximation due to the symmetry of the positive and negative maser velocities on the midline which preserves the product \(\sqrt{\tilde{M}}\,\sin{x}\). Thus, the degeneracy among the black hole and disk parameters cannot be broken unless an additional equation involving these parameters is available. 
For instance, several studies, including those performed by the MCP \cite{MCPXI,MCPXIII,Humphreys2013,Reid2019} implementing VLBI observations to combine megamaser positions, velocities and accelerations, 
estimate BH parameters with a Keplerian rotation curve model. The inclusion of maser accelerations allows for a different disentanglement of the $M$ and $D$ parameters. However, the orbital periods for megamaser systems are of the order of $10^{3}$ yr and require monitoring programs of the maser emission lasting several years. The warping maser disk offers an alternative way of decoupling $M$ and $D$ based on the geometry of accretion disks that, in principle, could render rough values for the $M$ and $D$ BH parameters hosted at the center of AGNs when precise measurement of the warping disk parameters and the peculiar reshift of the host galaxy is provided.

\section{Discussion and conclusions}
\label{sec:3}

The obtained closed formulas (\ref{eq:Dsolita})-(\ref{eq:Msolita}) constitute the main contribution of this work since, in principle, they allow the characterization of the principal parameters of a supermassive BH, expressing its mass and distance from Earth in the language of purely measurable astrophysical quantities. The decoupling of these quantities arises thanks to the addition of the simplest version of a disk warping to the relativistic model for the frequency shift given by Eqs. (\ref{eq:R})-(\ref{eq:x}). 

Notwithstanding,
it is worth noting that in this particular model, the expressions for $M$ and $D$ exhibit degeneracy in the warping parameters $\theta_1$ and $di/dr$, which limits their immediate applicability in astrophysical megamaser systems. Nevertheless, this modeling can be further developed to incorporate more realistic megamaser configurations where this degeneracy is not present, rendering models that could prove applicable in the near future.

These results could be relevant due to the fact that with this megamaser based method, the use of the scale ladder for distance measurements at cosmological scales is dispensed, implying the bypass of increasing uncertainties because only a single distance measurement to the galaxy hosting the BH is performed \cite{MCPXIII}. This is in contrast with methods employing standard candles, where strong assumptions about their cosmological environment are made and, therefore, big uncertainties must be handled.

On the other hand, we would like to finally remark that the mass determination of astrophysical AGN BHs is crucial for understanding their intrinsic physical properties as well as the evolution of their host galaxy. In this sense, it seems that supermassive BHs represent gravitational centers that become heavier in time in order to hold the structure of galaxies against the tendency of their dissolution caused by the expansion of the universe \cite{Farrah2023a,Farrah2023b}.

%
%

\begin{acknowledgements}
Authors are grateful to D. Villaraos, D.A. Martínez-Valera, U. Nucamendi and M. Momennia for fruitful discussions, 
to FORDECYT-PRONACES-CONACYT for support under grant No. CF-MG-2558591, to VIEP-BUAP as well as to SNII.  A.G.-J. acknowledges financial assistance from SECIHTI
through the postdoctoral grant No. 446473.
\end{acknowledgements}



\end{document}